\documentclass[conference]{IEEEtran}
\IEEEoverridecommandlockouts
\usepackage{cite}
\usepackage{amsmath,amssymb,amsfonts}
\usepackage{algorithmic}
\usepackage{graphicx}
\usepackage{textcomp}
\usepackage{xcolor}
\def\BibTeX{{\rm B\kern-.05em{\sc i\kern-.025em b}\kern-.08em
    T\kern-.1667em\lower.7ex\hbox{E}\kern-.125emX}}
\begin{document}

\title{UT-AISTimprt submission for ICME 2026 Grand Challenge on Academic Text-to-Music Generation}

\author{
\IEEEauthorblockN{
Shunsuke Yoshida\IEEEauthorrefmark{1},
Yu-Hua Chen\IEEEauthorrefmark{2},
Satoru Fukayama\IEEEauthorrefmark{2}
}
\IEEEauthorblockA{\IEEEauthorrefmark{1}University of Tokyo, Japan}
\IEEEauthorblockA{\IEEEauthorrefmark{2}National Institute of Advanced Industrial Science and Technology (AIST), Japan}
}

\maketitle

\begin{abstract}

This work investigates the effect of batch sampling strategies during training for text-to-audio music generation under low-data and small-scale model settings. 
This paper describes our approach and findings for the ICME 2026 Grand Challenge on Academic Text-to-Music Generation. 
Training data are clustered using either text embeddings or audio embeddings, and samples with similar characteristics are grouped within the same mini-batch to mitigate gradient interference.
The effects of modality and cluster granularity on clustering are analyzed.
Results show that clustering based on text embeddings achieves better performance on objective evaluation metrics than clustering based on audio embeddings.
In addition, different cluster granularity leads to different behaviors across evaluation criteria:
a moderate number of clusters performs best on objective metrics, while a larger number of clusters tends to exhibit music with more coherent structure in listening tests.

\end{abstract}

\begin{IEEEkeywords}
music generation, text-to-audio, batch sampling, clustering 
\end{IEEEkeywords}

\section{Introduction}
\label{sec:intro}
Machine learning for music generation often suffers from a limited amount of publicly available training data compared to text or image domains.
As a result, models for music generation are frequently trained on small-scale datasets, particularly in conditional and personalized generation settings. In practice, training small-scale models from scratch remains a realistic and commonly adopted setting.

With a limited amount of training data, using small-scale models can help mitigate overfitting, but their representational capacity is limited.
Consequently, the way training data are selected and presented during learning has a strong impact on generation performance.
Restricting training data to a narrow genre or condition often reduces the diversity of output.
On the other hand, naively mixing heterogeneous data can destabilize training of small-scale models due to gradient interference among diverse training data. The limited model capacity of the small-scale models have less flexibility to disentangle heterogeneous generative factors~\cite{gradientspace, gast}.

Similar issues have been reported in the natural language processing (NLP) community, particularly in the context of multi-task learning and instruction tuning~\cite{commonit,gast,gradientspace}. When multiple heterogeneous tasks or instructions are jointly learned, gradient conflicts can degrade both training efficiency and model performance. To address this problem, recent studies have proposed focusing on batch construction during training rather than on global data mixing ratios~\cite{commonit}. 
Commonality-aware Instruction Tuning (CommonIT) clusters training samples in advance and constructs each mini-batch from a single cluster, thereby increasing intra-batch homogeneity while preserving diversity across batches.

Inspired by CommonIT, this work investigates clustering-based batch sampling for text-to-audio music generation under low-resource and scratch-training conditions. The conditions are aligned with the ICME 2026 Grand Challenge on Academic Text-to-Music Generation~\cite{hsieh2026academic}. Since text-to-audio models involve both text and audio modalities, the choice of modality used to define similarity becomes a key decision. In this submission, training samples are clustered using embedding representations derived from either a text encoder or an audio encoder, and mini-batches are constructed from samples within the same cluster.

This work reports three main findings: 
\begin{itemize}
    \item clustering-based batch sampling improves the performance of small-scale music generation models under low-data settings,
    \item text-based clustering achieves better performance in objective evaluation metrics than audio-based clustering, and
    \item varying the number of clusters reveals a trade-off between objective evaluation metrics and perceptual quality. Objective metrics favor a moderate number of clusters, while a larger number of clusters tends to produce more coherent audio in qualitative listening.
\end{itemize} 

\section{Clustering-based batch sampling for text-to-audio music generation}
\label{sec:method}

\subsection{Overview}
This work builds a Text-to-Audio music generation system based on the official baseline provided for the ICME 2026 Grand Challenge on Academic Text-to-Music Generation. 

The system takes a text caption as input and generates a corresponding audio waveform through conditional audio generation.
This work investigates the batch construction strategies in training-time and keeps the model architecture unchanged. The original baseline architecture is preserved, while training mini-batches are constructed differently to analyze the impact of data sampling strategies under low-data and small-scale model settings.
The same model architecture and training configuration are used for both the Efficiency Track and the Performance Track, and identical models are submitted without modifications specific to each track.

\subsection{Baseline Text-to-Audio Model}
The method adopts the FluxAudio~\cite{FA} model implemented within the MeanAudio~\cite{MO} framework, which serves as the official baseline.
FluxAudio performs Text-to-Audio generation using a Flux-style Transformer trained with the Conditional Flow Matching objective.
The training data consists of the Jamendo dataset~\cite{jamendo}, processed through the official preprocessing pipeline. Each audio track is segmented into 10-second clips and encoded into latent representations using a variational autoencoder (VAE) paired with a BigVGAN vocoder.
For text conditioning, embeddings extracted from two pretrained encoders, T5 and CLAP, condition the generation of audio latent representations.

The model is trained from scratch. The architecture, loss function, and most of the training hyperparameters follow the official baseline configuration. Training uses the FluxAudio-L model variant without architectural modifications. Detailed specifications of the baseline model appear in the Grand Challenge workshop paper~\cite{hsieh2026academic} and the official repository.

\subsection{Clustering-based Batch Sampling}

\subsubsection{Clustering}

In our investigation, two types of embedding representations define data similarity for the clustering:

\begin{itemize}
    \item \textbf{Text-based clustering}: text embeddings extracted by a text encoder from caption text, and
    \item \textbf{Audio-based clustering}: 
    audio embeddings extracted by an audio encoder from audio signals
    latent representations obtained from encoded audio signals. 
\end{itemize}

Text-based clustering uses CLAP to extract text embeddings that capture semantic information related to music. 
Audio-based clustering relies on latent representations produced by the VAE encoder associated with the BigVGAN-based audio pipeline. 

For both text-based and audio-based clustering, k-means is used to partition the training data. We experiment with different numbers of clusters (K = 50 and 500), with K = 1 corresponding to the baseline configuration of the challenge.

\subsubsection{Batch Construction} 
For each mini-batch, batch construction draw samples exclusively from a single cluster. 
At the beginning of each epoch in training, a list of cluster index is randomly shuffled and a cluster index is selected sequentially by following the shuffled list. Samples from the selected cluster index creates several successive mini-batches until all of the samples in a cluster is sampled.
When a cluster does not contain enough samples to fill a complete batch, the remaining samples are skipped for the current epoch and deferred to the next cluster.

\subsection{Training and Inference}
Training largely follows the official baseline configuration, with adjustments made for computational constraints and training stability.
The batch size is set to 32, and training proceeds for a total of 600k optimization steps.
Optimization uses AdamW with a learning rate of $1 \times 10^{-4}$ and a linear warm-up period of $1000$ steps.
Gradient norm clipping with a threshold of 1.0 stabilizes optimization.
All models were trained using a single NVIDIA RTX 6000 Ada Generation GPU, except for the text-based clustering model with 500 clusters, which was trained using two GPUs due to higher computational cost.

During inference, the trained model receives a single text caption and generates a corresponding music. The clustering-based sampling strategy is used only during training and is disabled at inference time.
All inference-time parameters, including audio duration and sampling configuration, follow the official baseline setup.


\section{Evaluation}
\label{sec:eval}

\subsection{Official Challenge Evaluation}

\subsubsection{Evaluation Setup}

Two models were submitted for the official evaluation of the Grand Challenge, as permitted by the challenge regulations.
One model applies text-based clustering with the number of clusters set to $500$, while the other uses audio-based clustering.
Both models were trained according to the method described in Section~\ref{sec:method}.

The official evaluation follows the evaluation pipeline and metrics provided by the challenge organizers.

In Phase~1 of the challenge (Objective Evaluation), submissions are ranked based on a composite score derived from the following objective metrics:

\begin{itemize}
    \item \textbf{Fr\'echet Audio Distance (FAD).}
    This metric measures the distributional similarity between generated audio and a reference audio set hidden by the organizers. 
    Scores are computed using the CLAP-Laion-Music checkpoint
    \textit{music\_audioset\_epoch\_15\_esc\_90.14.pt} as the embedding model.

    \item \textbf{CLAP Score.}
    This metric evaluates semantic alignment between the input text prompt and the generated audio.
    Scores are computed using the same CLAP checkpoint as used for FAD.

    \item \textbf{Concept Coverage Score (CCS).}
    This metric measures how well multiple musical concepts contained in each prompt (e.g., tempo, instrumentation, genre, and mood) appear in the generated audio.
\end{itemize}

Evaluation uses publicly released text prompts, while the reference audio set used for FAD computation remains hidden.
Submitted audio must be at least 10 seconds long, and only the first 10 seconds of each audio sample are used for evaluation.

\subsubsection{Official Submission Results}
\label{sec:official-results}

\begin{table*}
    \centering
    \caption{Official evaluation results on the Grand Challenge test set. 
    Results of all non-proposed models are cited from the Grand Challenge workshop paper~\cite{hsieh2026academic}.
    }
    \begin{tabular}{lccccc}
        \hline
        \textbf{Model} & \textbf{params}  & \textbf{Train Data(hours)}& \textbf{FAD$\downarrow$} & \textbf{CLAP$\uparrow$} & \textbf{CSS$\uparrow$}\\
        \hline
        proposed (Text-500)        & 480M & 3.7K & 0.646 & 0.260 & 0.767\\
        FluxAudio-S (Organizer's Baseline) & 120M & 3.7K & 0.757 & 0.088 & 0.592\\
        Stable Audio Open\cite{SAO}       & 1.1B & 7.3K & 0.574 & 0.321 & 0.800 \\
        MusicGen-small\cite{MG}       & 300M & 20K & 0.574 & 0.370 & 0.875\\
        MusicGen-medium\cite{MG}      & 1.5B & 20K & {\bf 0.548} & 0.353 & {\bf 0.892}\\
        MusicGen-large\cite{MG}      & 3.3B & 20K & 0.553 & {\bf 0.379} & 0.888\\
        MeanAudio-S-Full\cite{MO}       & 120M & 10K & 0.649 & 0.210 & 0.808\\
        MeanAudio-L-Full\cite{MO}       & 480M & 10K & 0.660 & 0.202 & 0.783\\
        \hline
    \end{tabular}
    \label{tab:official-results}
\end{table*}

Table~\ref{tab:official-results} reports the official evaluation scores of the submitted model, the organizer-provided baseline, and several publicly available reference models. 
Direct comparisons across models are challenging due to differences in model size, training data, and experimental settings. Notably, our models are trained on the smallest training dataset, equivalent to that used for the official baseline model.
Nevertheless, the proposed model (Text-500) demonstrates competitive performance within this diverse set of models.
In particular, the proposed model achieves better scores than the organizer-provided baseline (FluxAudio-S) across all metrics.

Compared to models with a similar number of parameters, such as MeanAudio-L, the proposed approach achieves competitive performance despite using less than half the amount of training data. this result suggests that clustering enables more effective utilization of the available data.

\subsection{Additional Experimental Evaluation}
\label{sec:local-eval}

\subsubsection{Experimental Setup}
\label{sec:local-setup}

Additional experiments aside from the official evaluation were conducted for more fine-grained and controlled analysis.
All additional experiments share the same model architecture and training configuration described in Section~\ref{sec:method}, and differ only in the data sampling strategy.

The evaluation focuses on the following factors:

\begin{itemize}
    \item Differences in clustering modalities (text-based and audio-based)
    \item Differences in the number of clusters (1, 50, and 500)
\end{itemize}

Additional evaluation uses CLAP Score and FAD, consistent with the official evaluation. 
However, the additional evaluation results are solely for relative comparison between methods in the additional evaluation, since the reference dataset and audio preprocessing may not exactly match the official evaluation setup. 

The baseline model corresponds to the special case of clustering with a single cluster, which reduces to standard batch sampling without clustering. All models share the same architecture and training configuration, and differ only in the batch sampling strategy. 


\subsubsection{Additional Evaluation Results}
\label{sec:local-results}

\begin{table}[t]
    \caption{Additional evaluation results comparing clustering strategies and cluster sizes.}
    \label{tab:local-results}
    \centering
    \begin{tabular}{lcc}
        \hline
        \textbf{Model} & \textbf{FAD$\downarrow$} & \textbf{CLAP$\uparrow$}  \\
        \hline
        Baseline        & 0.503 & 0.200  \\
        Text-500         & 0.498 & 0.206 \\
        Text-50         & {\bf 0.491} & {\bf 0.217}  \\
        Audio-500       & 0.502 & 0.205  \\
        Audio-50        & 0.495 & 0.209  \\
        \hline
    \end{tabular}
\end{table}
Table~\ref{tab:local-results} present the results of the additional experiments. ``Text-$K$'' and ``Audio-$K$'' denote models trained with text-based and audio-based clustering using $K$ clusters, respectively. 
Overall, clustering-based methods outperform the baseline (no clustering) on both CLAP Score and FAD. Text-based clustering achieves higher CLAP scores than audio-based clustering for the same number of clusters. It also yields consistently lower FAD values. 

Number of clusters affects performance. Models with 50 clusters perform best on both metrics. Increasing the number of clusters to 500 slightly degrades performance. These results are consistent across both modalities.

\section{discussion}
\label{sec:discussion}

\subsection{Discussion on Official Evaluation Results}

The official challenge results show that the proposed method outperforms the model with the same amount of training data (3.7K hours for FluxAudio-S). Due to the differences in model size between the proposed method and the FluxAudio-S, the comparison needs further clarification by aligning the number of model parameters.   

The proposed method performs competitively against with models with similar parameter sizes (MeanAudio-S-Full and MeanAudio-L-Full), except for the MusicGen-small. The proposed method uses less than half the amount of training data compared to MeanAUdio-S-Full and MeanAudio-L-Full. These results suggest that the proposed method is more data-efficient.

\subsection{Discussion on Choice of Modality in Clustering}

Text-based clustering consistently outperforms audio-based clustering on objective metrics. This result indicates that grouping samples in the text embedding space is more effective for conditional music generation under low-resource settings.

A plausible explanation lies in how the condition for music generation aligns with the batch structure. 
Text-based clustering organizes training batches using the same modality (text) that conditions the generation process.
This alignment encourages more consistent gradients with respect to the prompt.
In contrast, audio-based clustering emphasizes acoustic similarity, which does not necessarily correspond to the semantic structure imposed by the text prompts.
For small models trained from scratch, this mismatch may degrade the effectiveness of audio-based clustering in stabilizing training.

Qualitative observations further support this explanation. 
In informal listening, audio generated by models trained with text-based clustering tended to better reflect prompt-level characteristics such as genre and overall mood.
By contrast, audio-based clustering occasionally emphasized consistency in timber without clearly reinforcing semantic intent.
Although no formal subjective evaluation was conducted, these observations are consistent with the quantitative trends.
Overall, these results suggest that, in conditional music generation with limited data and model capacity, clustering strategies that operate in the same semantic space as the condition offer a practical advantage.

\subsection{Discussion on Number of Clusters}
The number of clusters influences model performance. Models with 50 clusters outperform those with 500 clusters across both metrics. This trend holds across both text-based and audio-based clustering strategies. 

From an optimization perspective, adequate number of clusters may provide a favorable balance between intra-batch homogeneity and inter-batch diversity.
Using too few clusters reduces the method to standard batch sampling, allowing heterogeneous samples within a batch and potentially reintroducing gradient interference. 
Conversely, overly fine-grained clustering can limit batch diversity and decrease data utilization efficiency within each epoch, which may negatively affect objective metrics such as CLAP Score and FAD.

Interestingly, qualitative observations suggest a different tendency.
In informal listening, models with 500 clusters often produce audio with better prompt adherence and higher perceptual quality.
These samples tended to present more coherent musical structure and fewer perceptual artifacts across the generated music tracks.
While these models achieved lower scores on objective metrics, they were frequently perceived as producing more consistent and satisfying musical outputs.

This discrepancy highlights a potential gap between current objective evaluation metrics and perceptual aspects.
Objective metrics tend to emphasize distributional alignment and coverage,
whereas perceptual quality benefits from temporal coherence and structural consistency within each sample, which is encouraged by more fine-grained clusters.
The appropriate cluster granularity may therefore depend on the target objective, such as optimizing automated metrics or prioritizing perceptual coherence.

\subsection{Limitations and future works}

\subsubsection{Hyperparameter exploration}
This work evaluates representative configurations for the number of clusters (1, 50, and 500) and fixes the number of training iterations.
A comprehensive exploration of hyperparameters related to the number of clusters, training epochs, and batch size is beyond the scope of this study.
As a result, the reported findings primarily illustrate general trends and the effectiveness of clustering-based batch sampling, rather than optimal achievable performance.

\subsubsection{Multi-modal clustering}
The proposed approach applies clustering based on either text embeddings or audio embeddings, but not both simultaneously.
Integrating information from multiple modalities, for example through joint or multi-view clustering, may further improve training stability and generation quality.
Exploring such combined strategies remains an important direction for future work.

\subsubsection{Data augmentation}
This study does not systematically evaluate the interaction between data augmentation and clustering-based batch sampling.
While data augmentation is widely used to improve model generalization, applying clustering-based sampling may cause augmented samples to concentrate within the same cluster as their original audio clips.
In such cases, the diversity introduced by augmentation may not be fully exploited at the batch construction level.
Designing augmentation-aware clustering or sampling strategies represents an avenue for future investigation.

\subsubsection{Model and data scale}
The experiments focus on scratch training with relatively small models and limited training data.
The extent to which the proposed sampling strategy generalizes to larger model architectures or larger datasets remains unclear.
Evaluating scalability across different model sizes and data regimes constitutes an important future direction.

\section{conclusion}
\label{sec:conclusion}

This work investigated the impact of training-time batch construction strategies in text-to-audio music generation.
By clustering training data based on text and audio embedding representations and collecting similar samples within the same mini-batch, the effects of batch construction on training behavior and generation performance were analyzed under low-data and small-scale model settings.
The results indicate that clustering based on text embeddings achieves better performance on objective evaluation metrics than clustering based on audio embeddings.
With respect to the number of clusters, objective metrics favor a moderate cluster size, whereas informal listening suggests that models trained with a larger number of clusters can, in some cases, produce music that better follows the given text prompts and exhibits higher overall perceptual quality.
These findings indicate that, even under identical data and model architectures, batch construction plays an important role in shaping the behavior of text-to-audio music generation models.
Future work includes extending this approach to clustering methods that integrate both text and audio, as well as investigating batch design strategies that account for interactions with data augmentation.

\bibliographystyle{IEEEbib}
\bibliography{icme2026references}

\end{document}